\documentclass[9pt,twocolumn,twoside]{opticajnl}
\journal{opticajournal} 

\setboolean{shortarticle}{true}

\usepackage{siunitx}

\title{Demonstration of optical spring in an un-detuned cavity containing an optical parametric amplifier}

\author[1]{Jian Liu}
\author[1]{Juntao Pan}
\author[1]{Carl Blair}
\author[2]{Jue Zhang}
\author[3]{Hengxin Sun}
\author[1]{Li Ju}
\author[1,*]{Chunnong Zhao}

\affil[1]{ARC Centre of Excellence for Gravitational Wave Discovery, Department of Physics, University of Western Australia, 35 Stirling Highway, Crawley, WA 6009, Australia}
\affil[2]{ARC Centre of Excellence for Gravitational Wave Discovery, Centre of Gravitational Astrophysics, Australian National University, Canberra ACT 2601, Australia}
\affil[3]{State Key Laboratory of Quantum Optics and Quantum Optics Devices, Institute of Opto-Electronics, Shanxi University, Taiyuan 030006, China}
\affil[*]{chunnong.zhao@uwa.edu.au}

\begin{abstract}
Here we demonstrate the capacity to manipulate the optical spring (OS) effect by employing an optical parametric amplifier (OPA) within an optical cavity. We observed more than a factor of 2 increase in the OS frequency shift with the OPA. We also showed for the first time that the OS can be tuned by solely adjusting the OPA phase and showing an un-detuned cavity exhibiting an optical spring. The method can be applied to gravitational wave detectors in the signal recycling configuration to realize narrow bandwidth high sensitivity. The OS can be tuned to align the detector peak sensitivity frequency to known frequency continuous gravitational wave signals, dynamically tuned to track the gravitational wave signal from merging compact binaries or tuned to search for the post-merger signal of known binary coalescence.

\end{abstract}

\setboolean{displaycopyright}{false} 

\begin{document}

\maketitle

\section{Introduction}
Currently operating second-generation ground-based gravitational wave detectors (GWDs) including Advanced LIGO\cite{aasi2015advanced}, Advanced Virgo\cite{acernese2014advanced}, and KAGRA\cite{aso2013interferometer} are now being pushed towards the designed sensitivity where quantum noise is dominating the entire band from tens of hertz up to a few kilohertz. Though 'classical' ways such as increasing the laser power and increasing the test masses weight can shape the quantum noise limited sensitivity in the GWDs, the sensitivity remains limited by the free mass standard quantum limit (SQL). Various 'non-classical' quantum techniques have been proposed and implemented in GWDs to tackle quantum noise. A sensitivity enhancement of more than 6 dB below the shot noise limit has been demonstrated with external phase squeezed vacuum states injection\cite{acernese2019increasing, tse2019quantum}. Further with an external optical filter cavity or interferometer itself via EPR entanglement\cite{ma2017proposal} to rotate the squeezing angle depending on frequency, a technique called frequency-dependent squeezing\cite{mcculler2020frequency,acernese2023frequency}, the broadband quantum noise reduction has been achieved\cite{ganapathy2023broadband}. 

Alternatively, the optomechanical interactions between the optical field and suspended test masses inside the GWDs can correlate the phase and amplitude quadratures of the optical field, resulting in pondermotive squeezing that can greatly improve the detector sensitivity to below the SQL\cite{braginsky1999low}. This was analyzed in detail by Buonanno and Chen in the context of GWDs with a blue-detuned signal recycle cavity (SRC)\cite{buonanno2001quantum} where the detector can be considered an 'optical spring' detector where the free test mass is coupled with an optical spring. The coupled system can resonate at a pair of frequencies, namely the OS frequency and optical resonance frequency. At these frequencies, the detector quantum noise limited sensitivity is significantly enhanced. The OS frequency is normally around a few to tens of hertz, depending on the SRC detuning and the arm cavity power. Increasing the arm cavity power can shift the OS frequency higher but brings undesired effects\cite{zhao2006compensation,evans2015observation,liu2018angular}. It is also worthwhile to note that along with the extra optical rigidity, the optomechanical interaction also results in optical heating that excites the mechanical motion of the suspended test mass\cite{aspelmeyer2014cavity}, external feedback control or optical damping will be needed\cite{rehbein2008double,singh2016stable}.

The idea of optical-spring-enhanced detectors was further extended to the dynamic tuning of optical springs by adjusting the SRC parameters, allowing tracking of expected chirps of GW signals\cite{meers1993dynamically}, but due to complexity in controls, this was not yet employed in GWDs. Somiya $et~al.$ proposed a new scheme with OPA inside the SRC that allows shifting the OS higher without increasing the arm cavity power\cite{somiya2016parametric}. Korobko $ et~al.$ further found that the OPA phase was crucial for shaping the interferometer's response\cite{korobko2018engineering}. Recently, Somiya $et~al.$ expanded the idea in the context of a dual recycled interferometer without arm cavities, they found that by properly choosing parameters, the OS frequency can be shifted to kilohertz, allowing detections of GW signals from the postmerger remnant of binary neutron star collisions\cite{kentaro2023intracavity}. 
The experiment carried out by Zhang $et~al.$ confirmed that the OS can be enhanced by OPA in a detuned linear cavity with a strong carrier\cite{zhang2023optical}. In this letter, we demonstrated that both the OPA gain and phase can be used to manipulate the OS effect.

Our model is that of a linear cavity with a moveable end mirror and a nonlinear crystal inside. A 1064 nm laser beam together with a 532 nm beam is injected into the cavity, the infrared beam resonates inside the cavity while the green beam doesn't. By adopting Huang $et~al.$'s formalism\cite{Huang2009Enhance}, we can write the system Hamiltonian as
 \begin{equation}\label{eq:hami}
 \begin{aligned}
		H=&\hbar\left(\omega_{c}-\omega_{L}\right)\hat{c}^{\dagger}\hat{c}-\frac{\hbar \omega_c}{L}\hat{q} \hat{c}^{\dagger}\hat{c}+\frac{1}{2}\left(\frac{\hat{p}^{2}}{m}+m \omega_{m}^{2} \hat{q}^{2}\right)\\
   &+i \hbar \varepsilon \left(\hat{c}^{\dagger}-\hat{c}\right)+i \hbar G \left(e^{i \theta} \hat{c}^{\dagger2}-e^{-i \theta} \hat{c}^{2}\right),
   \end{aligned}
	\end{equation}
where $\omega_{L}$ and $\omega_{c}$ represent the frequency of the laser and the resonant frequency of the cavity, $\hat{c}$ and $\hat{c}^{\dagger}$ are the annihilation and creation operators of the cavity field; $\hat{p}$ and $\hat{q}$ represent the momentum and position operators for the mechanical resonator; $ \varepsilon =\sqrt{2\gamma}\sqrt{P /\left(\hbar \omega_{L}\right)}$ is the cavity field driving strength with input laser power $P$; $\gamma=\pi c/(2 \mathcal{F} L)$ is half-linewidth of the cavity with finesse $\mathcal{F}$ and cavity length $L$, $G$ is the non-linear gain of the OPA which is proportional to the green pump power, and $\theta$ is the phase of the pump field. 

We can then solve the equations of motion and obtain the steady-state values
\begin{equation}
p_{s}=0, q_{s}=\frac{\hbar\omega_c\left|c_{s}\right|^{2}}{m L  \omega_{m}^{2}},
c_s=\lvert c_s \rvert e^{i \phi}=\frac{\gamma- i\Delta+2 G e^{i \theta}}{\gamma^{2}+\Delta^{2}-4 G^{2}}\varepsilon ,
\label{eq:cavityfield}
\end{equation}
where
\begin{equation}		
\Delta=\Delta_{0}-\frac{\hbar \left(\omega_c\right)^{2}\left|c_{s}\right|^{2}}{m L^2 \omega_{m}^{2}},\phi=\arctan{\frac{2 G \sin[\theta]-\Delta}{\gamma+2 G \cos[\theta]}}
\end{equation}	
are the effective cavity detuning, and the phase of the cavity field, respectively. We can further solve the equations of motion in the frequency domain and get the modified mechanical susceptibility
\begin{equation}\label{eq:OPA:mechsus}
\chi_{xx}^{-1}=-m(\omega^2+i[\gamma_{\rm{m}}+\Gamma_{\rm{os}}]\omega-[\omega_{ \rm{m} }^2+K_{\rm{os}}/m]),
\end{equation}
where the optical spring constant $K_{\rm{os}}$ and optical damping $\Gamma_{\rm{os}}$ are defined as: 
\begin{equation}\label{eq:OPA:springconst}
\begin{aligned}
		K_{\rm{os}}&=Re[\chi_{\rm{os}}^{-1}]\\
		&= \frac{2 \hbar (\frac{\omega_c}{L})^2 \lvert c_s \rvert ^2(\gamma ^2+\Delta ^2-4 G^2-\omega^2)(\Delta+2 G\sin[\theta-2 \phi])}{(\gamma ^2+\Delta ^2-4 G^2-\omega^2)^2+4\gamma^2 \omega ^2},
\end{aligned}
\end{equation}
\begin{equation}\label{eq:OPA:damping}
\Gamma_{\rm{os}}=\frac{Im[\chi_{\rm{os}}^{-1}]}{m_{\rm{eff}}\omega}
            =\frac{4 \hbar (\frac{\omega_c}{L})^2 \lvert c_s \rvert ^2 \ \gamma (\Delta+2 G \sin[\theta-2 \phi])}{m_{\rm{eff}} ((\gamma ^2+\Delta ^2-4 G^2-\omega^2)^2+4\gamma^2 \omega ^2)}.
\end{equation}

It can be seen from Eq.(\ref{eq:OPA:springconst},\ref{eq:OPA:damping}) that the mechanical frequency $\omega_{m}$ and damping rate $\gamma_{m}$ are modified, by both the OPA gain and phase. The new effective mechanical frequency and damping rate are $\omega_{\rm{eff}} = \sqrt{\omega_{ \rm{m} }^2+K_{\rm{os}}/m}$ and $\gamma_{\rm{eff}} = \gamma_{\rm{m}}+\Gamma_{\rm{os}}$. Without OPA, the results reduce to the normal optomechanical cavity case\cite{aspelmeyer2014cavity}. A more detailed calculation can be found in supplementary \ref{supp:theory}.

\section{Experiment setup}

\begin{figure}[ht]
\centering
\includegraphics[width=1\linewidth]{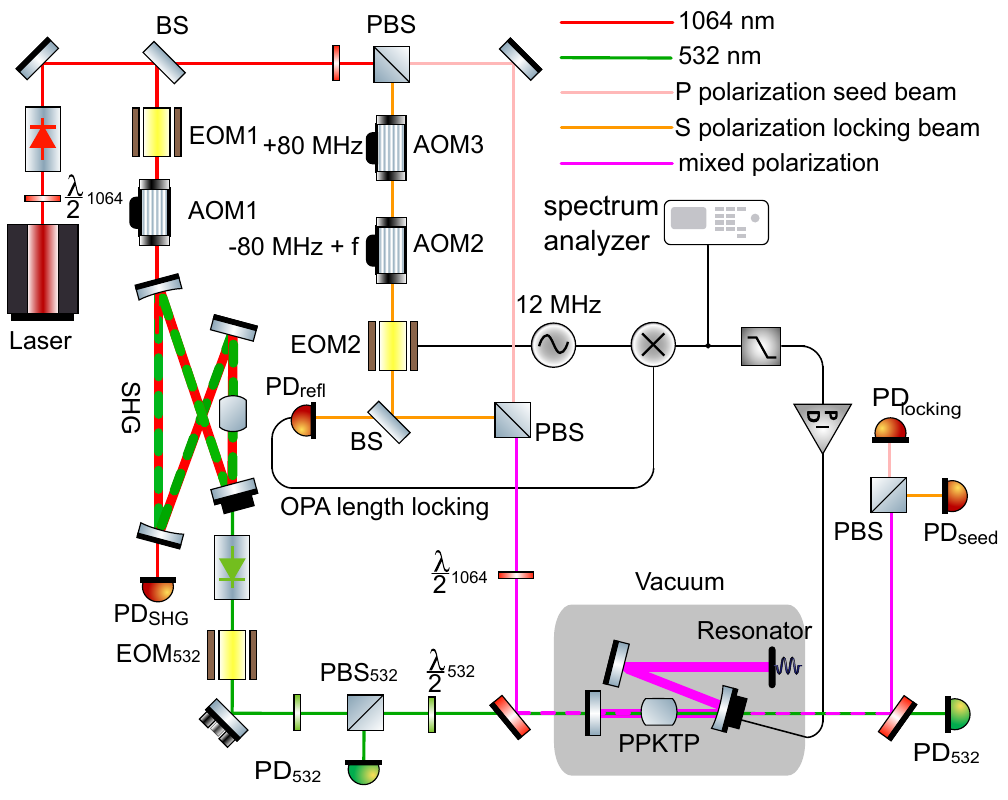}
\caption{Schematic of the experiment. 1064 nm EOMs are driven at 12 MHz and 532 nm EOM is driven at 300 kHz. All three beams have independent power control allowing arbitrary relative power injection and they are also independently monitored at the transmission port.}
\label{fig:schematic}
\end{figure}

The schematic of the experiment is presented in Fig.\ref{fig:schematic}. A 2 W Nd:YAG laser produces a linear polarized 1064 nm beam and it is split into two parts. Half of the beam goes to a second harmonic generator (SHG). The SHG is locked via the standard PDH locking technique\cite{black2001introduction} and can produce a 532 nm green beam with a power of about 500 mW. The other half of the 1064 beam is further split via a polarized beam splitter (PBS) into two paths: the S-polarized locking beam and the P-polarized seed beam.

The locking beam goes through two acoustic-optical modulators (AOM) to shift the frequency relative to the seed beam for cavity detuning purposes. It combines with the seed beam at a second PBS. A half-wave plate is located after this PBS to adjust the beam polarization. The 1064 nm infrared beams are further combined with the S-polarized 532 nm green beam at a dichroic mirror that is HR-coated for 1064 nm and AR-coated for 532 nm. Both the seed beam and locking beam are mode-matched to the cavity and the green beam is focused down to below 50 um in radius to maximize the intensity on the crystal. The whole cavity sits in a high vacuum chamber that has a pressure of about 1e-5 mbar to reduce the air-damping.

The cavity is a Z-shape linear cavity with three half-inch mirrors and a micro-resonator as the end mirror. The total cavity length is 260 mm with a 50-um waist on the input mirror and a second 150-um waist on the resonator. The input mirror is AR coated at 1064 nm and 532 nm at the outside surface while the inner surface is AR coated for 532 nm but HR coated with 99.8$\%$ power reflectivity for 1064 nm. A 1$\times$2$\times$10 mm PPKTP crystal is placed right after the input mirror, and both the 1$\times$2 mm surfaces are AR coated with power reflectivity less than 0.1$\%$. The second cavity mirror is a concave mirror with a radius of curvature of 100 mm and the third one is a flat mirror. Both of them have a power reflectivity of 99.98$\%$ at 1064 nm and less than 5$\%$ at 532 nm. By reaching the micro-resonator, the green power would be attenuated by more than three orders of magnitude, so that the thermal effect induced by the green beam absorption on the resonator is negligible. The transmitted beams from the curved mirror are split via a dichroic mirror and then through a PBS depending on wavelength and polarization, then reach photodiodes for power monitoring.

The resonator is made of a 420 $\mu$m thick 20 mm wide square silicon frame that is etched with multiple through holes of different sizes leaving windows of a multi-layer AlGaAs/GaAs coating bonded on one side. Each of the coating windows can be used as an individual resonator with different effective masses, resonant frequencies, and quality factors. These windows are located within the central 7 mm $\times$ 7 mm area of the frame. To minimize the loss from clamping, the whole frame is clamped using a 9 mm radius ring shape holder. The resonator used in this experiment is 0.64 mm $\times$ 0.64 mm, the drum shape mode frequency is 85 kHz with an effective mass of 1.75 $\mu$g and a quality factor of about 2500 measured in vacuum (see details in supplementary \ref{supp:mechaincalresonator}). The cavity linewidth of the seed beam was measured to be 1.14 MHz, corresponding to a cavity finesse of around $\mathcal{F} = 500 $ for the seed beam. 




\section{Results}

We first measured the optical spring effect without the green pump beam injection. Only a 3 mW locking beam together with a 10 mW seed beam is injected into the cavity. The cavity is stably locked to the locking beam and the seed beam is detuned from the cavity resonance by adjusting the driving frequency of AOM2. The OPA cavity length locking bandwidth is about 3 kHz and the loop gain around the mechanical mode frequency is about -80 dB, so the loop should have little effect on the mechanical mode. The mode frequency and linewidth can be obtained through the OPA cavity length locking error signal via a spectrum analyzer. 

\begin{figure}[ht]
\centering
\includegraphics[width=0.8\linewidth]{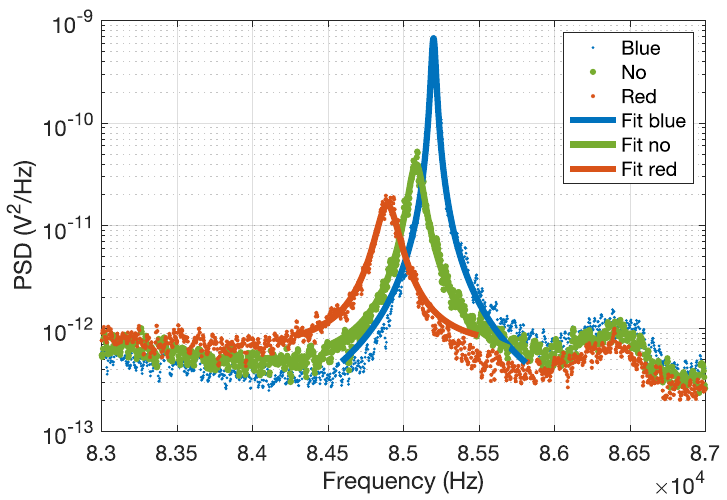}
\caption{Examples of the 85 kHz drum mode power spectrum density under blue, red, and no detuning and the corresponding Lorenzian fittings.}
\label{fig:OSdataexample}
\end{figure}

\begin{figure}[ht]
\centering
\includegraphics[width=0.8\linewidth]{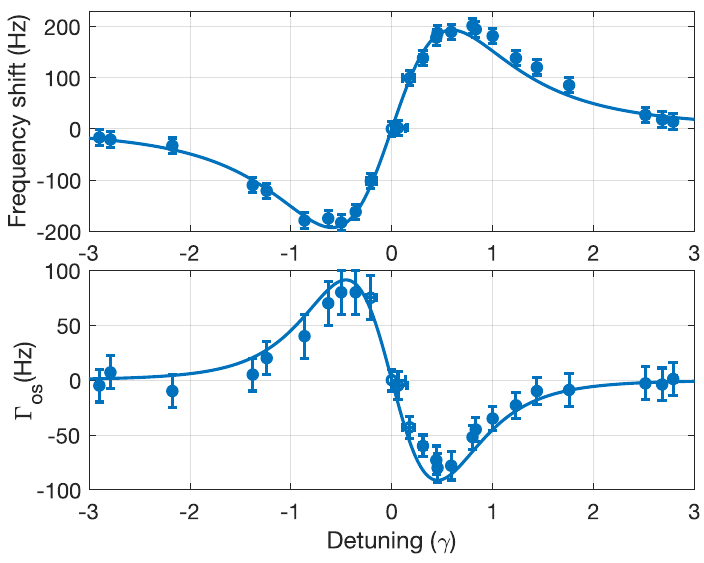}
\caption{Optical spring effect without OPA. The seed beam power is 10 mW and the transmission PD reading is 2.6 V. }
\label{fig:OSwithoutOPA}
\end{figure}

Examples of the measurement of mechanical mode power spectrum density from a spectrum analyzer are shown in Fig.\ref{fig:OSdataexample}. Such curves are measured over a range of cavity detuning of [$-3~\gamma$, $-3~\gamma$]. Each of the measured curves is then fitted with a Lorentzian shape to obtain the mode central frequency and the linewidth. The OS effect without OPA is summarized in  Fig.\ref{fig:OSwithoutOPA}. The maximum frequency shift is about $\pm$200 Hz and the maximum damping or anti-damping is about $\pm$100 Hz at around $\pm 0.25~\gamma$ detuning. The measurements agree with the theory quite well. In the theory curve, we have taken into account the 20\% power loss through mode mismatch, the loss from the vacuum tank window, and 0.65 for the overlap factor between the cavity fundamental mode and the resonator drum mode. Each measurement curve takes about 20 seconds to average the noise. During the whole measurement, the natural frequency of the mode can shift tens of hertz due to the thermal effect and the mode linewidth can increase due to pressure change in the vacuum tank. The detuning has to be set to zero after a few measurements to allow calibration of the mode's natural frequency and linewidth to reduce errors in the OS results. 

In the following step, we then injected about 50 mW p-polarized green pump laser into the cavity, the pump laser is phase modulated by driving a 532 nm EOM at 300 kHz. The PPKTP crystal temperature is stabilized at 34 degrees to meet the phase-matching condition of the OPA to maximize the down-conversion process. The seed laser 1064 nm beam will be amplified inside the cavity and is monitored through transmission. The transmission signal is then mixed with the 300 kHz local oscillator to generate the phase-locking error signal. Depending on the relative phase between the pump beam and the seed beam, the seed beam power can be either amplified or de-amplified. We define the OPA power gain as
\begin{equation}	g_{\rm{OPA}}=\left|\frac{c_s(G)|_{\Delta\rightarrow0,\theta\rightarrow0}}{c_s(G=0)|_{\Delta\rightarrow0,\theta\rightarrow0}}\right|^2 = \left(\frac{\gamma}{\gamma - 2G}\right)^2.
\label{eq:opa:G}
\end{equation}

Without the green pump injection, 0.5 mW seed laser registered 0.2 V in the transmission PD, while with 50 mW pump beam and OPA phase locked in amplification, the transmission read 2.0 V, meaning that the $g_{\rm OPA}$ = 10, corresponding to a non-linear gain $G$= 0.2 MHz.

\begin{figure}[ht]
\centering
\includegraphics[width=0.8\linewidth]{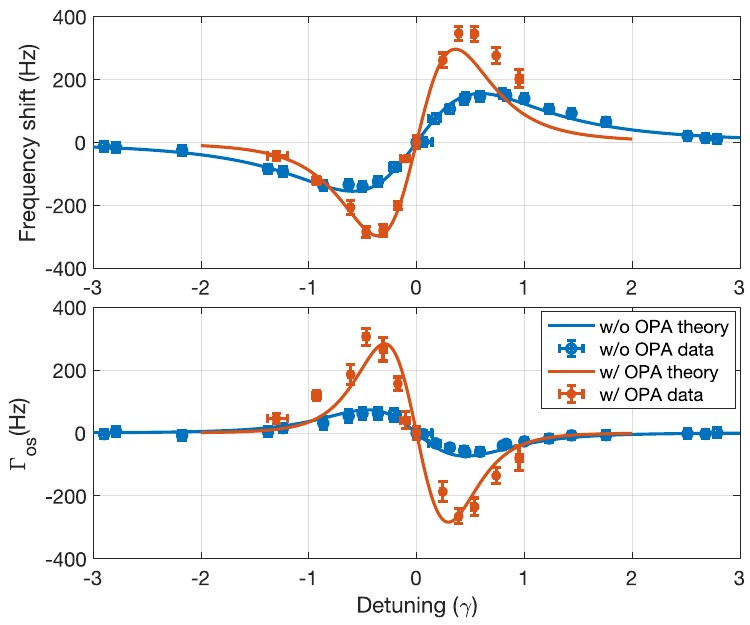}
\caption{OS as a function of the cavity detuning with and without OPA. Note that the x-axis $\gamma$ refers to the original cavity half-linewidth without OPA. The effective cavity linewidth is less with OPA. The w/o OPA traces are from Fig.\ref{fig:OSwithoutOPA} by normalizing the intra-cavity power from 2.6 V to 2.0 V on transmission PD.}
\label{fig:OSenhancedOPA}
\end{figure}

We then repeated the OS measurements while keeping the pump beam power stable and the OPA phase locked to maximum amplification. By adjusting the AOM2 driving frequency, we were able to detune the cavity and obtain the OPA-enhanced OS results as shown in Fig.\ref{fig:OSenhancedOPA}. The peak frequency shift was enhanced by a factor of 2 and the damping(or anti-damping) was enhanced by a factor of 4. 

\begin{figure}[ht]
\centering
\includegraphics[width=0.8\linewidth]{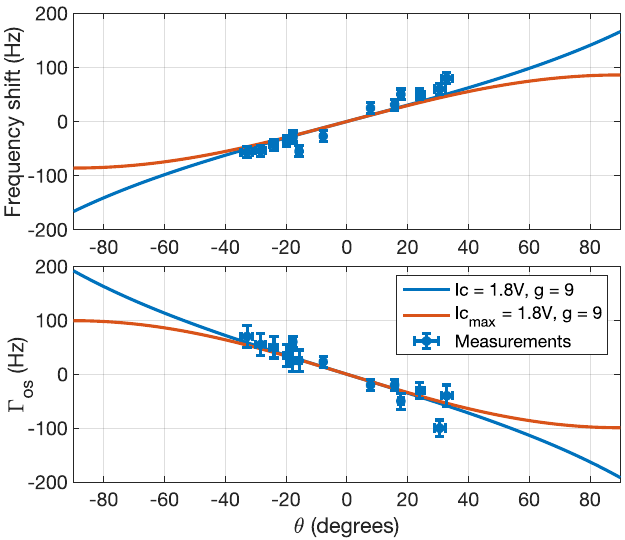}
\caption{OS as a function of the OPA phase $\theta$. The blue trace has the intra-cavity power normalized to 1.8 V while the orange trace assumes the maximum power is 1.8 V and the intra-cavity power decreases while the phase changes. It can be seen that when the phase is small, the OS effects are the same for the two traces.}
\label{fig:Phasetuning}
\end{figure}

As shown in Eq.(\ref{eq:OPA:springconst},\ref{eq:OPA:damping}), the OPA enhancement is not only achieved by tuning the OPA gain $G$ but also the OPA phase $\theta$. In our next measurement, we kept the $\Delta = 0$ by adjusting the AOM2. The OPA phase, instead of being locked on $\theta = 0$ that has maximum amplification, is locked at an arbitrary value by adding a phase offset to the phase-locking loop error point. By comparing the offset with the full size of the error signal, the phase can be calculated (See supplementary \ref{supp:pllerrorsig}). The intra-cavity power would change while tuning the OPA phase and ideally, the input power needs to be adjusted to maintain the intra-cavity power the same for a fair comparison. But within a range from $-40^{\circ}$ to $40^{\circ}$, the intra-cavity power can be considered stable. The measurements are shown in Fig.\ref{fig:Phasetuning}. 

\section{Conclusions}
We have experimentally demonstrated that the OS effect can be enhanced by an OPA inside the cavity by more than a factor of two. Both the OPA gain and phase can be tuned independently or together to modify the OS effect. This method can be applied in large-scale gravitational wave detectors to improve detector sensitivity by shifting the OS frequency from a few hertz where low-frequency seismic noise dominates, to a few hundred hertz where the quantum noise dominates. The peak sensitivity frequency can be set to a certain value for monitoring gravitational wave signals that have known frequency or dynamically tuning the peak sensitivity frequency would allow real-time tracking of the gravitational events. 












\begin{backmatter}
\bmsection{Funding} This work was supported by the Australian Research Council

\bmsection{Acknowledgment} 
The authors would like to thank
\smallskip

\bmsection{Disclosures} The authors declare no conflicts of interest.

\bmsection{Data availability} Data underlying the results presented in this paper are available from the corresponding author upon reasonable request.

\bmsection{Supplemental document}
See Supplements for supporting content. 

\end{backmatter}

\bibliography{OS_OPA}

\bibliographyfullrefs{sample}

\section{Supplementary materials}
\subsection{Theoretical model}\label{supp:theory}
The system Hamiltonian can be written as
 \begin{equation}\label{eq:hami}
 \begin{aligned}
		H=&\hbar\left(\omega_{c}-\omega_{L}\right)\hat{c}^{\dagger}\hat{c}-\frac{\hbar \omega_c}{L}\hat{q} \hat{c}^{\dagger}\hat{c}+\frac{1}{2}\left(\frac{\hat{p}^{2}}{m}+m \omega_{m}^{2} \hat{q}^{2}\right)\\
   &+i \hbar \varepsilon \left(\hat{c}^{\dagger}-\hat{c}\right)+i \hbar G \left(e^{i \theta} \hat{c}^{\dagger2}-e^{-i \theta} \hat{c}^{2}\right),
   \end{aligned}
	\end{equation}
with each term corresponding to the intra-cavity optical field, the interaction between the intra-cavity field with the mechanical mode, the energy of the mechanical mode, the driving from the injection, and the down-conversion process in the crystal, respectively.

The equation of motion can be expressed as
\begin{equation} \label{eq:OPA:eom}
\begin{array}{l}
\dot{\hat{q}}=\hat{p} / m \\
\dot{\hat{p}}=-m \omega_{m}^{2} \hat{q}+\hbar \frac{\omega_c}{L} \hat{c}^{\dagger}\hat{c}-\gamma_{m} \hat{p}+\hat{\xi} \\
\dot{\hat{c}}=i\left(\omega_{L}-\omega_{c}\right) \hat{c}+i  \frac{\omega_c}{L} \hat{q} \hat{c}+\varepsilon+2 G e^{i \theta} \hat{c}^{\dagger}-\gamma \hat{c}+\sqrt{2\gamma} \hat{c}_{\rm{in}}\\
\dot{\hat{c}}^{\dagger}=i\left(\omega_{c} - \omega_{L}\right) \hat{c}^{\dagger}-i \frac{\omega_c}{L} \hat{q} \hat{c}+\varepsilon+2 G e^{-i \theta} \hat{c}-\gamma \hat{c}^{\dagger}+\sqrt{2\gamma} \hat{c}_{\rm{in}}^{\dagger}.
\end{array}
\end{equation}
Here we included $\hat{\xi}$ as the Brownian noise operator and $\hat{c}_{\rm{in}}$ as the input vacuum noise operator both have a mean value of zero. The steady-state solution of the equation would be 
\begin{equation}
p_{s}=0, q_{s}=\frac{\hbar\omega_c\left|c_{s}\right|^{2}}{m L  \omega_{m}^{2}},
c_s=\lvert c_s \rvert e^{i \phi}=\frac{\gamma- i\Delta+2 G e^{i \theta}}{\gamma^{2}+\Delta^{2}-4 G^{2}}\varepsilon ,
\end{equation}
where 
\begin{equation}		
\Delta=\Delta_{0}-\frac{\hbar \left(\omega_c\right)^{2}\left|c_{s}\right|^{2}}{m L^2 \omega_{m}^{2}}, \phi=\arctan{\frac{2 G \sin[\theta]-\Delta}{\gamma+2 G \cos[\theta]}}
\end{equation}
are the effective cavity detuning, and the phase of the cavity field.
After linearisation, $\hat{q}\rightarrow q_{s}+\hat{q}$, $\hat{p}\rightarrow p_{s}+\hat{p}$, $\hat{c}\rightarrow c_{s}+\hat{c}$, and Fourier transformation of Eq.~\ref{eq:OPA:eom}, we can solve the equations in frequency domain and derive the position fluctuations of the movable mirror as
\begin{equation}
    \begin{aligned}
    \hat{q}(\omega)=&-\frac{1}{d(\omega)}\{(\Delta^{2}+(\gamma-i \omega)^{2}-4 G^{2})\hat{\xi}(\omega)\\
		   -&{i\hbar\sqrt{2\gamma} \frac{\omega_0}{L_{\rm{f}}}[((\omega+i \gamma-\Delta) \hat{c}_{s}+2 i G e^{i \theta} \hat{c}_{s}^{\dagger}) \hat{c}_{i n}^{\dagger}(\omega)+\rm{H.c.}]\}},
   \end{aligned}
	\end{equation}	
with
 \begin{equation}
        \begin{aligned}
		    d(\omega)=&2 \hbar 
		    (\frac{\omega_c}{L})^{2} \lvert c_s \rvert ^2 \left(\Delta+2 G \sin[\theta-2 \phi]\right)\\
            &+m\left(\omega^{2}-\omega_{m}^{2}+i \omega \gamma_{m}\right)\left[\Delta^{2}+(\gamma-i \omega)^{2}-4 G^{2}\right].
        \end{aligned}
    \end{equation}
The susceptibility of the mechanical resonator is modified by the optomechanical interaction as
	\begin{equation}
		\begin{aligned}
			&\chi_{xx}^{-1}=-\frac{d(\omega)}{\Delta^{2}+(\gamma-i \omega)^{2}-4 G^{2}}
			=\chi_{0}^{-1}+\chi_{\rm{os}}^{-1}\\
		    &=-m (\omega^2 - \omega_{\rm{m}}^2 + i\omega\gamma_{\rm{m}})
		    + \frac{2 \hbar(\frac{\omega_c}{L})^{2} \lvert c_s \rvert ^2\left(\Delta+2 G \sin[\theta-2 \phi]\right)} {\Delta^2+(\gamma - i\omega)^2 - 4 G^2},
		\end{aligned}
	\end{equation}
where $\chi_{0}=-\frac{1}{m (\omega^2 - \omega_{\rm{m}}^2 + i\omega\gamma_{\rm{m}})}$ is the intrinsic mechanical susceptibility and the second term $\chi_{\rm{os}}^{-1}$ represents the optomechanical modification to the susceptibility with the OPA. 
	To make it clear, we rewrite this equation as
 \begin{equation}
\chi_{xx}^{-1}=-m(\omega^2+i[\gamma_{\rm{m}}-\Gamma_{os}]\omega-[\omega_{ \rm{m} }^2+K_{\rm{os}}/m]),
\end{equation}
where the optical spring constant $K_{\rm{os}}$ and optical damping $\Gamma_{\rm{os}}$ are defined as: 
\begin{equation}
\begin{aligned}
		K_{\rm{os}}&=Re[\chi_{\rm{os}}^{-1}]\\
		&= \frac{2 \hbar (\frac{\omega_c}{L})^2 \lvert c_s \rvert ^2(\gamma ^2+\Delta ^2-4 G^2-\omega^2)(\Delta+2 G\sin[\theta-2 \phi])}{(\gamma ^2+\Delta ^2-4 G^2-\omega^2)^2+4\gamma^2 \omega ^2},\\
  \Gamma_{\rm{os}}&=\frac{Im[\chi_{\rm{os}}^{-1}]}{m_{\rm{eff}}\omega}
            =\frac{4 \hbar (\frac{\omega_c}{L})^2 \lvert c_s \rvert ^2 \ \gamma (\Delta+2 G \sin[\theta-2 \phi])}{m_{eff} ((\gamma ^2+\Delta ^2-4 G^2-\omega^2)^2+4\gamma^2 \omega ^2)}
\end{aligned}
\end{equation}
The new effective mechanical frequency and damping rate are $\omega_{\rm{eff}} = \sqrt{\omega_{ \rm{m} }^2+K_{\rm{os}}/m}$ and $\gamma_{\rm{eff}} = \gamma_{\rm{m}}+\Gamma_{\rm{os}}$.

\subsection{Mechanical resonator}\label{supp:mechaincalresonator}
\begin{figure}[ht]
    \centering
    \includegraphics[width=0.7\linewidth]{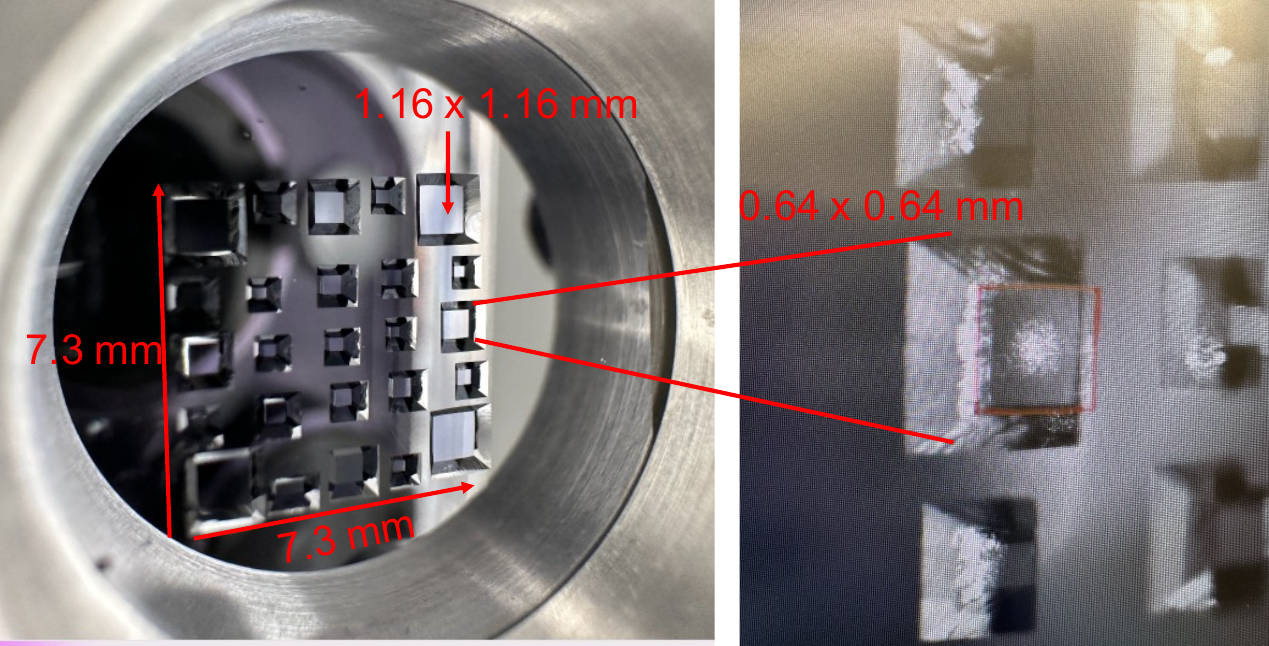}
    \caption{The mechanical resonator. Left penal: front view of the resonator, showing 25 different size windows. right panel: photo showing the laser beam on the 0.64 $\times$ 0.64 mm window, the beam size is about 150 $\mu$m.}
    \label{fig:resonator}
\end{figure}

The mechanical resonator used in the experiment is shown in Fig.\ref{fig:resonator}. It is a through-etched silicon frame with a multi-layer AlGaAs/GaAs coating bonded at the backside. There are in total 25 square shape windows with the biggest being 1.16 $\times$ 1.16 mm and the smallest being 0.175 $\times$ 0.175 mm. Each window is an independent resonator with different resonance frequencies and quality factors. The one presented in this work is of size 0.64 $\times$ 0.64 mm.

\begin{figure}[ht]
\centering
\includegraphics[width=0.7\linewidth]{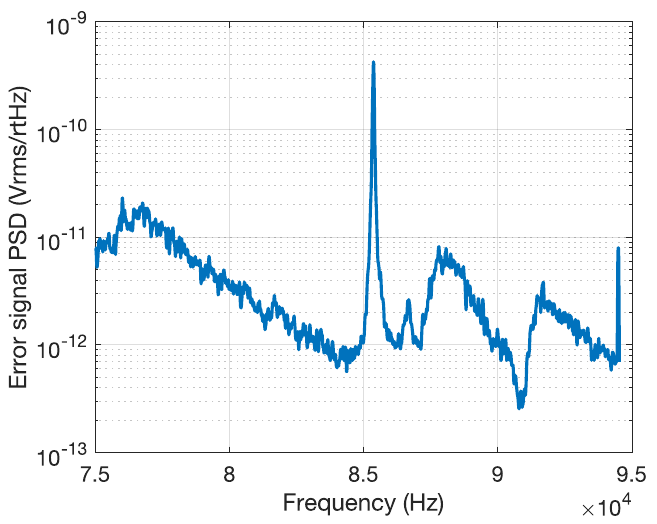}
\caption{Cavity locking error signal PSD, showing the mechanical mode at a frequency around 85 kHz. }
\label{fig:modestructure}
\end{figure}

The target mechanical mode frequency is about 85 kHz, with a quality factor of around 2500. Multiple modes associated with the whole frame or the frame holder have much lower quality factors, shown in Fig.\ref{fig:modestructure}.

\subsection{Phase locking loop error signal}\label{supp:pllerrorsig}
\begin{figure}[ht]
\centering
\includegraphics[width=0.8\linewidth]{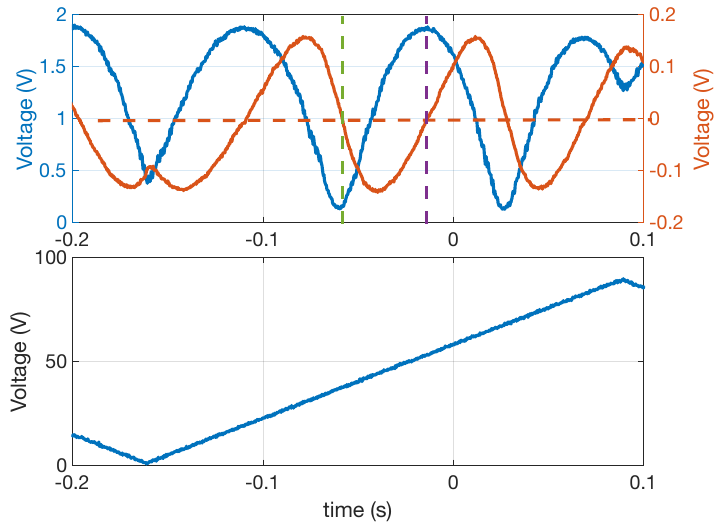}
\caption{Time traces of OPA phase scanning, the blue trace in the upper panel represents the power of the laser beam at the seed beam polarisation; the orange trace is the phase-locking loop error signal. The orange horizontal dashed line is the zero level of the phase-locking error signal. The vertical dashed lines indicate that the OPA is either at the maximum amplification (purple) or maximum de-amplification (green). The lower panel is the 2 Hz triangle scanning voltage applied on the PZT mirror where the phase locking loop error signal will be applied to lock the OPA phase.}
\label{fig:pllerrorsig}
\end{figure}

\begin{figure}
    \centering
    \includegraphics[width=1\linewidth]{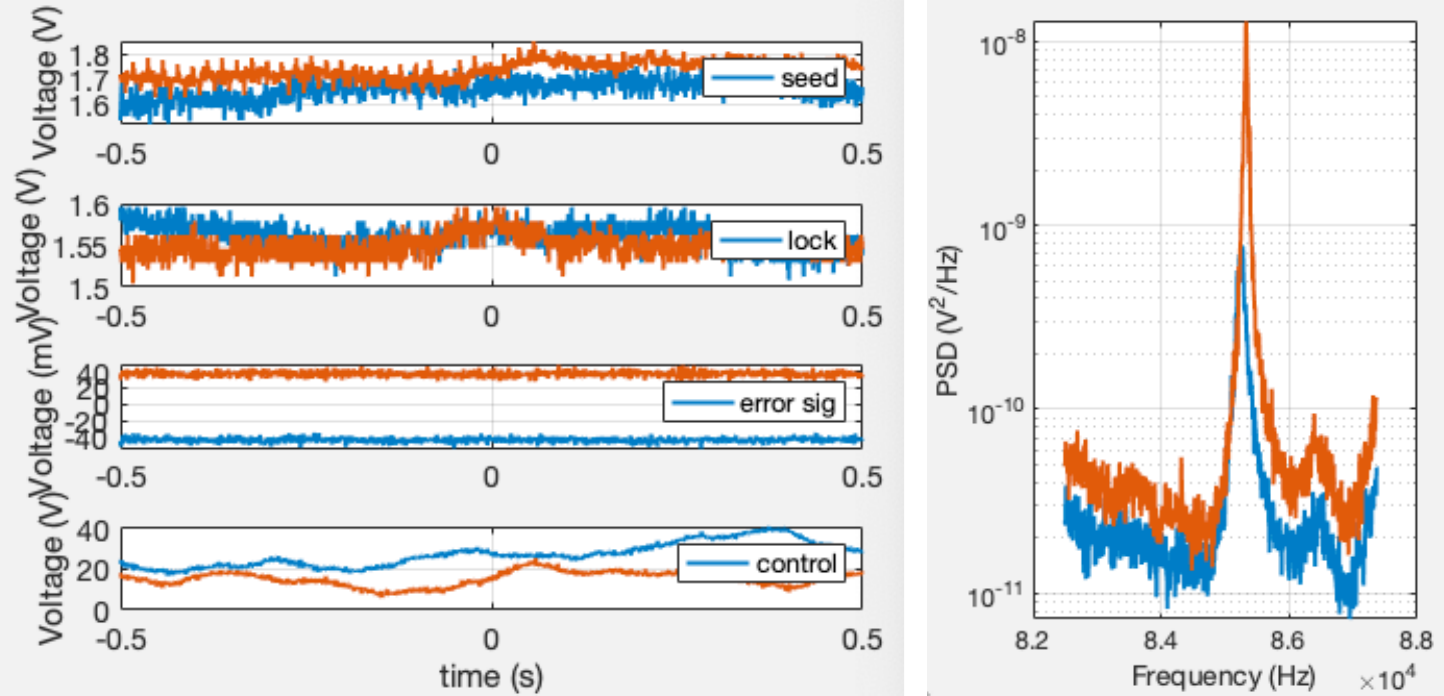}
    \caption{Optical spring effect by tuning OPA phase. The left panel from top to bottom shows the power of the beam at the seed beam polarisation, the power of the locking beam, the phase locking loop error signal, and the OPA phase locking loop feedback control signal. The blue traces are when the phase locking loop locked at the error signal equals -40 mV and 40 mV for the orange trace. The right side panel is the corresponding mechanical mode power spectrum measured from the OPA cavity length locking loop error signal, showing that the OS effect can be controlled by adjusting the OPA phase without cavity detuning.}
    \label{fig:opaphasetuning}
\end{figure}

The phase-locking loop error signal is obtained by modulating the green laser beam phase through an EOM at 300 kHz while monitoring the intra-cavity power in the seed beam polarization. The intra-cavity power would also be modulated because the OPA amplification is different for different OPA phases. The intra-cavity field is then transmitted through the cavity mirror and detected by the PD outside the vacuum chamber. The PD signal is then mixed with the 300 kHz local oscillator and low-passed to produce the phase-locking error signal. Fig.\ref{fig:pllerrorsig} top panel shows the time traces of the transmitted power in blue and the corresponding error signal in orange. When the OPA is locked at the error signal equal to zero, the OPA is either at maximum amplification or maximum de-amplification status. The OPA can also be locked at an arbitrary phase by offsetting the error point of the phase-locking loop. The error signal is a sinusoidal function of the OPA phase, thus by comparing the error signal offset ($e.g.$ 40 mV shown in Fig.\ref{fig:opaphasetuning}) and the full-size error signal (147.5 mV $\pm$ 7.5 mV), the OPA phase can be calculated ($\theta_{\rm{errsig = 40 mV}} ={15.7^\circ} ^{+0.9^\circ}_{-0.8^\circ}$ ).

 

\end{document}